\documentclass[onecolumn,showpacs,aps]{revtex4}
\usepackage{graphicx}
\begin{document}
\preprint{J. Chem. Phys. 128, 164706 (2008)}
\title{Bias-driven local density of states alterations and transport in ballistic molecular devices}
\author{Ioannis Deretzis}
\email{ioannis.deretzis@imm.cnr.it}
\author{Antonino La Magna}
\email{antonino.lamagna@imm.cnr.it}

\affiliation{Istituto per la Microelettronica e Microsistemi (CNR-IMM)\\
Stradale Primosole 50, Catania 95121, Italy}

\date{\today}

\begin{abstract}
We study dynamic nonequilibrium electron charging phenomena in ballistic molecular devices at room temperature that compromise their response to bias, and whose nature is evidently distinguishable from static Schottky-type potential barriers. Using various metallic/semiconducting carbon nanotubes and alkane dithiol molecules as active parts of a molecular bridge, we perform self-consistent quantum transport calculations under the non-equilibrium Green's function formalism coupled to a three-dimensional Poisson solver for a mutual description of chemistry and electrostatics. Our results sketch a particular tracking relationship between the device's local density of states and the contact electrochemical potentials that can effectively condition the conduction process by altering the electronic structure of the molecular system. Such change is unassociated to electronic/phononic scattering effects while its extent is highly correlated to the conducting character of the system, giving rise to an increase of the intrinsic resistance of molecules with a semiconducting character and a symmetric mass-center disposition.
\end{abstract}

\maketitle

\section{Introduction}
It is a common perception nowadays that both industry and academic research have largely focused their attention on molecular structures as active part candidates of nanoscale electronic devices. Such trend is other than unfounded, as reduced dimensionality could significantly advance miniaturization and improve performance of plausible logic and storage devices. Aside the increasing electronic transport measurement paradigms of synthetic organic/inorganic molecular structures\cite{Aren07}, it is undoubtable that carbon nanotubes (CNTs) have constituted the central point of molecular research during the last decade\cite{Anan06}. Moreover, recent zeolite template growth techniques\cite{Hulm03} have clearly demonstrated that the era of thin ($d\approx{}0.4nm$ diameter) CNTs cannot be considered too distant. Nonetheless advances in nanoscale engineering have brought knowledge of molecular electronics up to a level of integrated circuit assembling\cite{Chen06}, the comprehension of the quantum effects that rule the conduction mechanism has not been fully mastered, in particular when it comes to interface interaction issues between device and electrode atoms. Freshman electronics teaches that when two surfaces with different work functions come to contact, charge transfer takes place that tends to minimize this difference. Such effect has been repeatedly noted in the case of CNTs embedded to metallic electrodes\cite{Hein02,Chen05}, with Schottky-type potential barriers that vary on the basis of the metallic contact element and the CNT diameter. Although conceptually convenient, the categorization of the quality of metallic contacts considering only metal-CNT work function differences finds experimental inconsistency, e.g. in the case of group 10 transition metals palladium and platinum\cite{Jave03,Mann03}. Theoretically, the electronic configuration regime of a molecule changes with respect to its isolated equilibrium picture when a) the device is attached to metallic contacts (appearance of metal-induced states\cite{Pomo04}, broadening of energy levels\cite{Zahi03}, charge transfer due to work function differences\cite{Xue04}) and b) when the device goes out of equilibrium (bias-induced charge transfer\cite{Dere07}). It is therefore reasonable that a complete theoretical investigation of quantum transport effects in nanodevices accounts also for dynamic nonequilibrium processes\cite{Zahi05,Pomo04} apart from the static equilibrium ones. In such context we have preliminarily evidenced the presence of transmission probability alterations in ballistic semiconducting carbon nanotube molecular bridges\cite{Dere07}, purely related to the application of a bias on the two electrodes of the system. Moreover such effect manifests in a strong coupling regime at room temperature, excluding Kondo or Coulomb blockade implications. By presenting equilibrium/nonequilibrium local density of states (LDOS) distributions and electronic density modifications, this article attempts to intensely examine the physical processes that lead to the aforementioned behavior for various metallic/semiconducting CNTs, and generalize findings by demonstrating similar conduction aspects also in alkane dithiol molecules.

Modeling of carbon-nanotube-based devices can often result puzzling, since too short CNTs manifest contact-induced states that can significantly change the density of states spectrum with respect to the respective bulk device\cite{Pomo04}, while long `real-size' systems are computationally prohibited out of the single-orbital tight-binding model. This work implements a mathematical formalism that can capture geometrical, chemical and charging interactions without at the same time resulting computationally weighty. In this sense the non-equilibrium Green's function formalism has been based on an all-valence-electron extended H\"{u}ckel Hamiltonian\cite{Hoff63}, iteratively coupled to a three-dimensional Poisson solver in the self-consistent field regime. Such approximation can combine a) the flexibility of semiempirical methods, b) the qualitative aspects of the Extended H\"{u}ckel method in the calculation of CNT band structures\cite{Kien06}, c) the inclusion of charging effects under bias\cite{Zahi05} and d) the uniform description of both device and contacts, allowing at the same time multiple hundreds-of-atoms simulations with a relatively low CPU load. Based on this model, this study is focused on the correlation between the electrochemical potentials of the metallic electrodes and the local density of states of a biased molecular system, the change in the electronic density that such interaction provokes and the consequences of these processes on the conduction mechanism.

This paper is organized as follows: Section II describes the details of the mathematical/computational formalism, section III shows transmission probability and LDOS spectra for CNT and alkane dithiol systems out of equilibrium, section IV demonstrates changes in electronic density and potential profiles for a metallic and a semiconducting CNT, and finally in section V we discuss the presented results.

\section{Methodology}
The theoretical foundation of our computational model is based on the method developed by Zahid \textit{et al.}\cite{Zahi05}, which couples the Green's function formalism based on an extended H\"{u}ckel Hamiltonian with a complete neglect of differential overlap (CNDO) mean field theory for the inclusion of Coulomb interactions, with appropriate in-house mathematical enhancements that render computations more affordable. The basic points of such formalism are briefly described in the following: a non-equilibrium Green's function\cite{Datt95} is used for the quantum calculation of transport for various finite size CNTs and alkane dithiol molecules embedded between two semi-infinite metallic planes (source and drain contacts). Such approach is based on the single particle retarded Green's function matrix $G = [ES - H - \Sigma_L - \Sigma_R]^{-1}$, where $E$ is the scalar energy, $H$ is the device Hamiltonian matrix in an appropriate basis set, $S$ is the overlap matrix in that basis set and $\Sigma_{L,R}$ is the self energy, which includes the effect of scattering due to the left $(L)$ and right $(R)$ contacts. Self energy $\Sigma_{L,R}$ terms can be expressed as $\Sigma = ({\tau}-ES)g_s({\tau}^\dagger-ES)$, where $g_s$ is the surface Green function specific to the contact type and $\tau$ is the Hamiltonian relative to the mutual interaction between the device and the contact\cite{Datt95}. Device and contacts Hamiltonian matrices are obtained using an all-valence-orbital H\"{u}ckel method\cite{Hoff63}, where orbitals are approximated by Slater-type functions and the parameterization of diagonal elements is based on theoretical and experimental data of ionization energies for the respective atoms\cite{Zahi03,Dere06}. Non-diagonal Hamiltonian elements are calculated on the basis of the overlap matrices between different atomic orbitals\cite{Dere06}. Considering that the computational framework takes place in the coherent limit, a Landauer-type expression can be used for the current calculation:
\begin{equation} \label{curr}
I=\frac{2e}{h}\int_{-\infty}^{+\infty}dET(E)[f(E,\mu_L)-f(E,\mu_R)].
\end{equation}
In this expression $T(E)=Tr[{\Gamma_L}G{\Gamma_R}G^\dagger]$ represents the transmission probability for the device current as a function of energy and $\Gamma_{L,R}=i[\Sigma_{L,R}-\Sigma_{L,R}^\dagger]$ are the anti-Hermitian parts of the self-energy terms. The statistical Fermi-Dirac distribution of electrons $f(E,\mu_{L,R})$ in the contact at chemical potential $\mu_{L,R}$ has a particular importance here since simulation temperatures have been set to 300 K. Charging effects can be introduced in the model with the inclusion of a self-consistent potential $U_{SC}(\Delta\rho)$ that is a functional of the device density matrix difference between equilibrium and non-equilibrium states ($\Delta\rho=\rho-\rho_{eq}$), and which is added to the bare device's Hamiltonian ($H = H_{0} + U_{SC}(\Delta\rho)$). The $U_{SC}(\Delta\rho)$ is calculated self-consistently by the addition of three separate constituent terms\cite{Zahi05}:
\begin{equation}
\label{Usc}
U_{SC}(\Delta\rho)=U_{Laplace}+U_{Poisson}(\Delta\rho)+U_{Image}(\Delta\rho),
\end{equation}
whereas $\rho$ is given by
\begin{equation}
\rho=\frac{1}{2\pi}\int_{-\infty}^{+\infty}dE[f(E,\mu_L)G\Gamma_{L}G^\dagger+f(E,\mu_R)G\Gamma_{R}G^\dagger].
\end{equation}
$U_{Laplace}$ and $U_{Image}$ parts of the potential have been numerically calculated with a three-dimensional finite element method solving the $\nabla^{2}U=0$ equation with appropriate box boundary conditions\footnote{For the Laplace term: $U_{Laplace}=-qV_{S}$ at the source and $U_{Laplace}=-qV_{D}$ at the drain. For the Image term: $U_{Image}=-U_{Poisson}(\vec r)$ at the position of the device contacts.}. The calculation of the Poisson term can derive in the framework of the CNDO theory, using only the Hartree potential for the Coulomb interaction\cite{Zahi05}. CNDO parameters have been determined for device atoms using ionization potential and electron affinity data from ref.\cite{Hinz62}. All potential components have been evaluated on the atoms (a linear interpolation between grid and atomic sites has been implemented in the Laplace and Image term case).

For the integration of the density matrix at each step of the self-consistent process a contour integration technique has been applied\cite{Bran02} up to energies where the Fermi-Dirac coefficient is still a unity ($f(E,\mu_{L,R})\approx{}1$), in order to avoid poles on the complex energy plane for energies close to the system's Fermi level. Both real and complex parts of the density integral have been accurately evaluated using Gaussian quadrature formulas\cite{Pres92}. In the iterative procedure of potential calculation, apart from simple mixing techniques, Anderson mixing algorithms\cite{Eyer96} have been introduced for a fast albeit precise convergence. Finally parallelization techniques have been implemented in the programming code for a better exploitation of multiple computer CPUs.

\section{Local density of states and bias correlation in carbon nanotube and alkane dithiol molecules}
One of the most appealing aspects of semiempirical methods is their flexibility in the determination of various system parameters without necessarily compromising on the description of the qualitative characteristics of the physical system. In this sense the implemented method allows for a separate study of the effects of electron charging on the conduction mechanism under equilibrium and non-equilibrium conditions. Although there are plenty of theoretical works that affront the problem of CNTs and static Schottky potential barriers in a multidisciplinary level\cite{Xue04,Shan04,Pala07}, to the best of our knowledge there has been no systematic attempt to evidence charging manifestation under nonequilibrium conditions and its role on the propagation of current in CNT systems. In the current study, the nonequilibrium transport features have been separated from the electrostatic effects of equilibrium (Schottky barrier formations) by setting the Fermi level of the whole system (device and contacts) to the Fermi energy level of the respective bulk CNTs. Such condition imposes the alignment between the metallic contact's work function and the charge neutrality level of the CNT by shifting the energy bands of the device with respect to the ones of the contact, affecting only equilibrium charge transfer features while leaving all chemical and nonequilibrium interactions unchanged. In the laboratory, an analogous phenomenon can be realized by using a gate electrode strongly coupled to the device levels, to unidirectionally shift the CNT bands towards higher or lower energies, depending on the spectral position of the electrodes' work function.

\begin{figure}
	\centering
		\includegraphics[width=0.5\columnwidth]{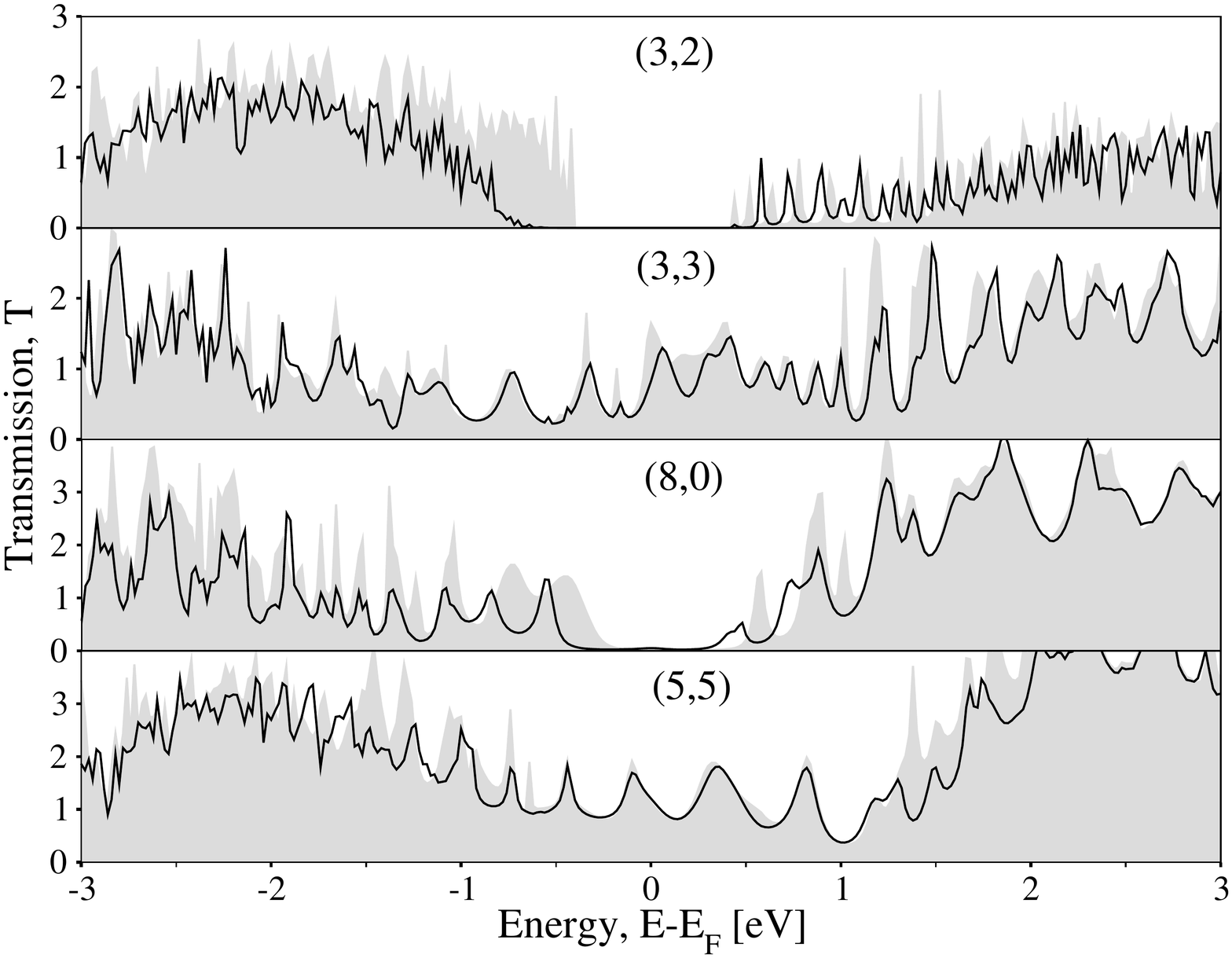}
	\caption{Transmission as a function of energy for a) a 4-unit-cell (3,2), b) a 15-unit-cell (3,3), c) an 8-unit-cell (8,0) and d) a 12-unit-cell (5,5) carbon nanotube system, for 0V (shaded area) and 2V (black line) applied bias. The contact metal is Au(111) whereas zero energy refers to the charge neutrality level of each CNT.}
	\label{fig:figure1}
\end{figure}

\begin{figure}
	\centering
		\includegraphics[width=0.5\columnwidth]{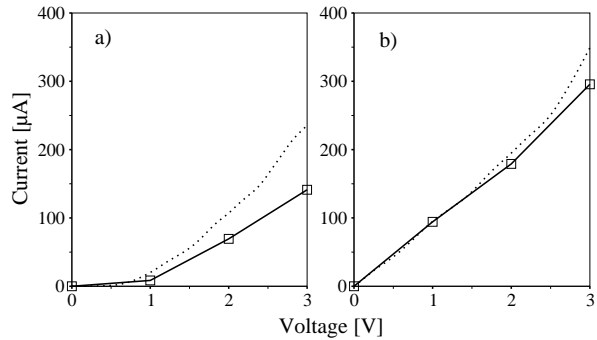}
	\caption{Current-Voltage curves for a) an 8-unit-cell (8,0) and b) a 12-unit-cell (5,5) carbon nanotube system. Solid lines represent the current value based on self-consistent calculations, whereas dotted lines are obtained without self-consistency, based on the simple potential profile of references \cite{Zahi03,Dere06}, where the potential fully drops in the interface area between the device and the contacts, while throughout the device it remains flat.}
	\label{fig:figure2}
\end{figure}

Strong interface coupling conditions have been explicitly imposed at this study by bringing Au(111) metallic electrodes at an $L=0.1nm$ distance from the two open edges of the CNTs\cite{Dere06}. Simulations have been undertaken for various metallic and semiconducting nanotubes in equilibrium and under bias, and figure \ref{fig:figure1} shows transmission function outcomes for CNTs with basis vector indices (3,2), (3,3), (8,0) and (5,5). The diameters of these CNTs are $d_{(3,2)}\approx{}0.34nm$, $d_{(3,3)}\approx{}0.4nm$, $d_{(8,0)}\approx{}0.63nm$, $d_{(5,5)}\approx{}0.68nm$, their lengths $L_{(3,2)}\approx{}7.5nm$, $L_{(3,3)}\approx{}3.6nm$, $L_{(8,0)}\approx{}3.4nm$, $L_{(5,5)}\approx{}2.9nm$, while their conducting character is fully determined by their helicity and (in the case of narrow CNTs) $\sigma-\pi$ hybrid orbital mixing\cite{Dere06}. The point of differentiation between semiconducting and metallic carbon nanotubes lies in the response of the intrinsic conducting capability of each system to bias. Semiconductor tubes, apart from shifting their transmission peaks corresponding to a shift of energy eigenvalues\cite{Zahi05}, demonstrate also a clear diminishment of their total transmission probability, which also affects the energy zone close to the Fermi level of the system. In terms of the Landauer formalism this can be translated in a decrease of the device's current-carrying capacity\cite{Dere07} and could be seen as an increase of the nanotube resistance compared to the one estimated from the equilibrium transmission spectrum(see figure \ref{fig:figure2}), which has no relation with phononic or electronic scattering interactions, static Shottky-type potential barriers or low-temperature/weak coupling effects. Rather, such process implies an implicit alteration of the propagation modes of current under non-equilibrium. On the other hand, although the phenomenon is present also for the metallic CNTs, its strength is evidently limited and the transmission coefficient does not diverge significantly from the equilibrium case. Such non-uniform situation cannot be simply attributed to energy level shifts and needs to be better evaluated.

\begin{figure}
	\centering
		\includegraphics[width=0.5\columnwidth]{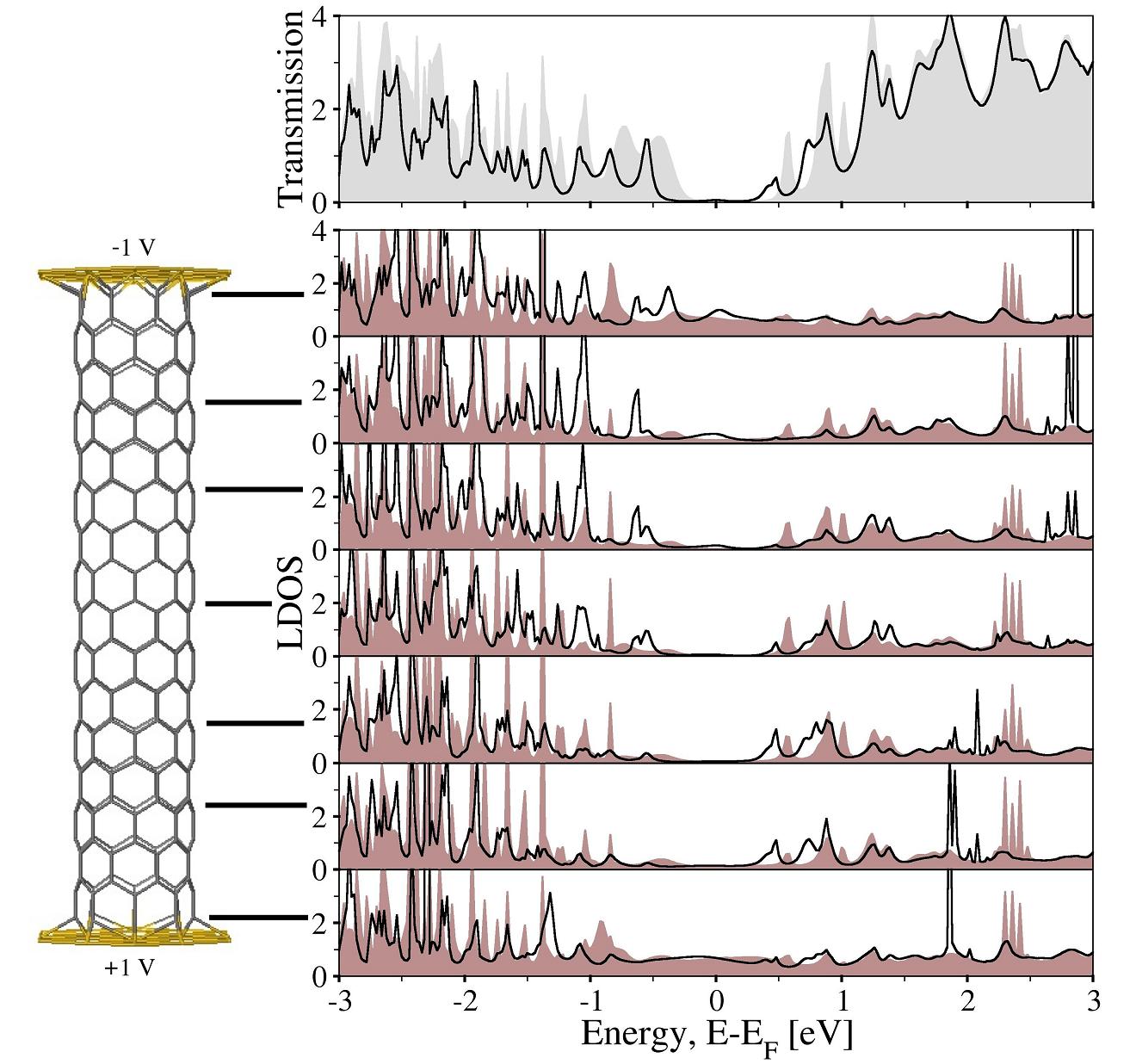}
	\caption{Geometry, transmission coefficient and ring LDOS evaluation as a function of energy for an 8-unit-cell (8,0) CNT for 0V (shaded area) and 2V (black line) applied bias.}
	\label{fig:figure3}
\end{figure}

\begin{figure}
	\centering
		\includegraphics[width=0.5\columnwidth]{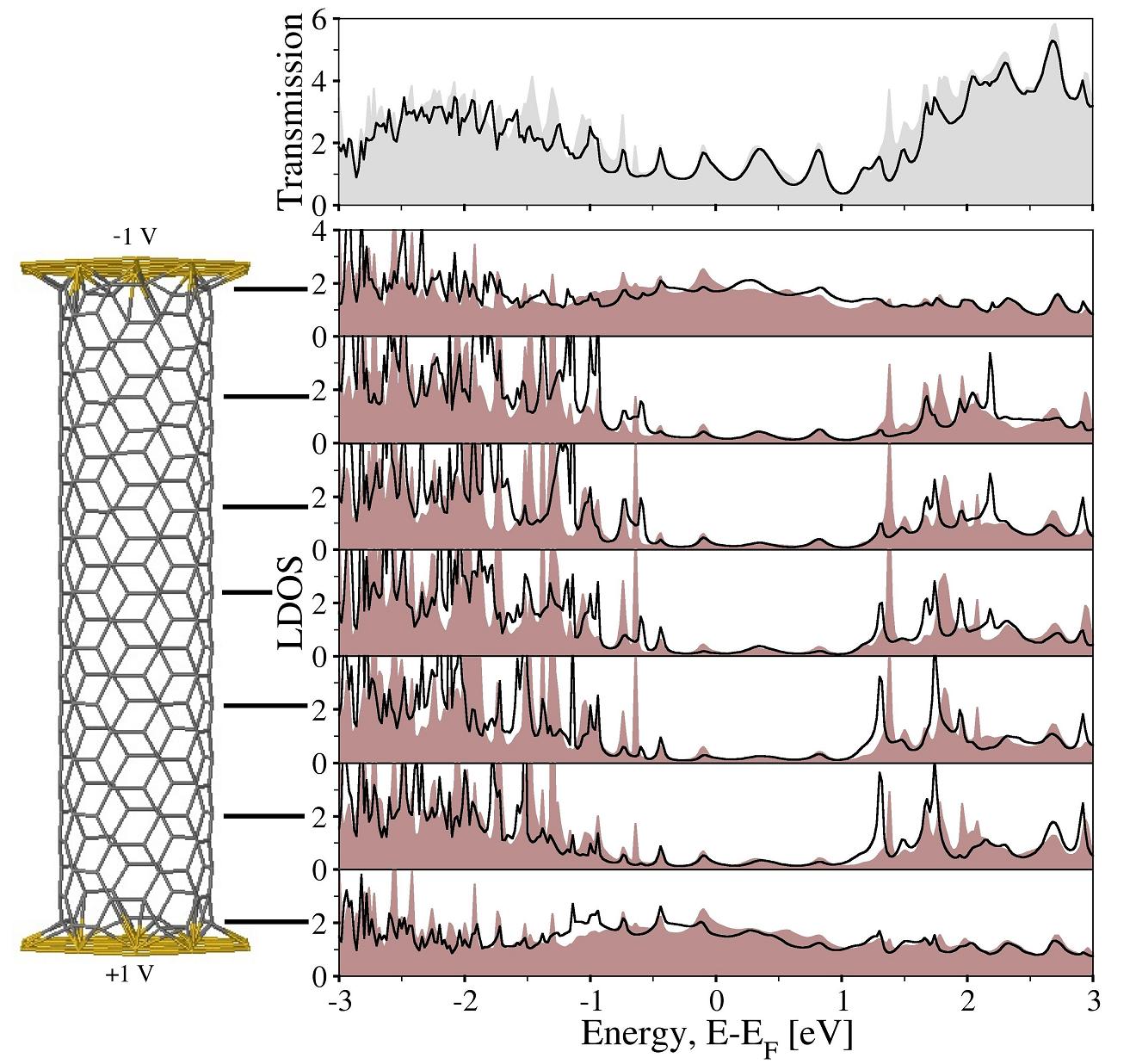}
	\caption{Geometry, transmission coefficient and ring LDOS evaluation as a function of energy for a 12-unit-cell (5,5) CNT for 0V (shaded area) and 2V (black line) applied bias.}
	\label{fig:figure4}
\end{figure}

The flow of current in molecules strongly depends on their electronic structure and its correlation with the metallic electrodes, in terms of interface chemical bonding and electrochemical potential positioning\cite{Zahi03}. Both qualitative and quantitative aspects of the latter are impressed on the device's density of states distribution. Hence, an LDOS ring analysis\cite{Trio05} has been performed for the (8,0) and (5,5) tubes in the cases of 0V and 2V applied bias, in order to evidence the processes that render the aforementioned metallic/semiconducting behaviors divergent. Figures \ref{fig:figure3} and \ref{fig:figure4} show the respective results. The observation of the distribution of states in the designated rings under equilibrium can evidence a high electronic density near the Fermi energy level for both CNTs close to the contacts that can be attributed to metal induced states, which arise from the tails of metallic electron wavefunctions that decay exponentially into the device body\cite{Xue04}. Moving towards the inner part of the CNT, such concentration tends to diminish to a finite value for the (5,5) tube and almost vanishes for the (8,0) tube, revealing respectively the metallic and semiconducting character of the studied CNTs. The application of bias has a qualitatively similar influence on the electronic bands of the two systems; nonetheless their different conducting character, the LDOS's of the various rings of the two systems tend to follow the movement of the electrochemical potential of the contact to which they are better correlated, both topologically and chemically. Thus, rings closer to the positively biased electrode (bottom electrode in figures \ref{fig:figure3},\ref{fig:figure4}) tend to shift their LDOS's towards lower energies, rings that are near to the negatively biased electrode (upper one in figures \ref{fig:figure3},\ref{fig:figure4}) tend to move LDOS's towards higher energies, while middle rings of both systems have LDOS's with minor differences that are attributed to geometrically -and therefore chemically- unequal couplings between the two electrodes and the device. Note that such minor mid-device alterations would not have been evidenced with a tight-binding representation of the metallic contacts that neglects geometrical differentiations in the reconstruction of interface chemical bonds\cite{Dere06}. The transmission probability of the semiconducting CNT is affected by the aforementioned LDOS redistribution in a bigger extent with respect to that of the metallic one, since this bidirectional LDOS motion at various parts of the CNT inserts and pulls out states from the conduction gap. These newly inserted gap states cannot always energetically correlate with states of neighboring rings. Such lack of correlation leads to a spatial localization of the electronic wavefunctions related to the aforementioned states (e.g. see the peaks appearing near the Fermi energy level in the non-equilibrium LDOS spectra of figure \ref{fig:figure3}) that cannot give any contribution to the globally evaluated transmission coefficient, and thus, the conductivity of the system. The implicit increase of the device resistance can therefore be seen as a bias-induced change of its electronic structure that provokes a decrease of availability of non-localized states within the conduction window. On the other hand, the magnitude of this effect is minor in the metallic CNT since shifted states can almost always correlate with neighboring ring states throughout the energy spectrum of interest for the conduction of current. In this case, the respective bias-induced change of the device's electronic structure does not lead to an extended localization of electronic wavefunctions.

\begin{figure}
	\centering
		\includegraphics[width=0.5\columnwidth]{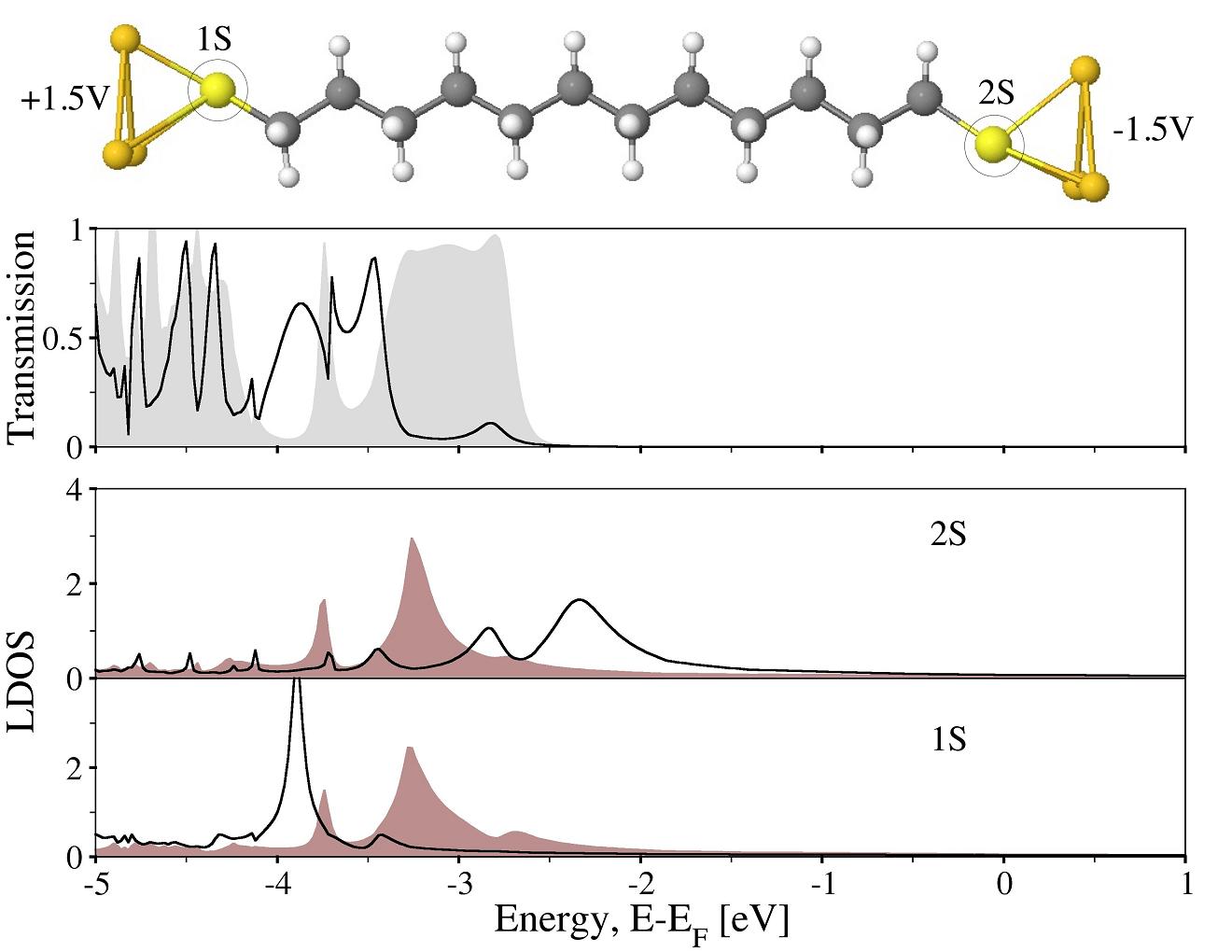}
	\caption{Geometry, transmission coefficient and sulfur atoms LDOS evaluation as a function of energy for a 12C alkane dithiol molecule for 0V (shaded area) and 3V (black line) applied bias.}
	\label{fig:figure5}
\end{figure}

The previous concepts can be better visualized in the analysis of few level molecular systems. In this sense alkane dithiols are theoretically ideal case study objects due to their small size and large conduction gap\cite{Aren07}. As such, the same type of investigation has been carried out for C8, C10 and C12 alkane dithiol molecules whereas representative results shown in figure \ref{fig:figure5} are for the C12 molecule. Here, in accordance with \textit{ab initio} reconstructions that calculate equilibrium Au-S distance between Au and $SCH_3$ complexes at $L_{Au-S}=0.25nm$\cite{Gonz07}, the vertical distance between the device and contacts has been set to $L=0.19nm$. The local density of states has been calculated on the atomic sites and fig. \ref{fig:figure5} demonstrates LDOS's for the two external sulfur atoms that spectrally give the most important density contribution with respect to C and H atoms. The LDOS trend that tracks electrochemical potentials is clearly confirmed also here, whereas states from the right sulfur atom that enter the conduction gap become localized and do not contribute to the transmission probability of the system. Attempting therefore a generalization of the presented results, it would be appropriate to state that the increase of the intrinsic resistance due to bias-induced factors should be expected in all molecules with a semiconducting/isolating character that exhibit a spatial symmetry with regard to their center of mass. The prerequisite of symmetry has a sense of isotropic contribution of the various parts of the molecule to the total electronic density. Asymmetry could have an impact similar to disorder on the electronic structure of a molecular system whereas the application of bias could affect singularly each distinct structure. In this sense, even if asymmetric molecule LDOS's are also expected to behave in an analogous to their symmetric counterparts way, their resistance may increase, decrease or remain unaffected.

The qualitative aspects of LDOS redistribution reported in this paragraph should be evidenced within the full range of semiemprirical/first-principles descriptions of a biased molecular system with a proper treatment of electron-electron interactions, even if the level of accuracy is expected to vary from case to case. On the other hand though, the intrinsic shortcomings of each method should influence on the correctness of the presented results.  In this work for example, a tight-binding representation of the device and the contacts could result inappropriate for small diameter CNTs\cite{Dere06}or alkane dithiol molecules, and lacking of realism at the contact-device interface schematization\cite{Dere06}.

\section{Spatial distribution of nonequilibrium charges on molecular devices}
\begin{figure}
	\centering
		\includegraphics[width=0.5\columnwidth]{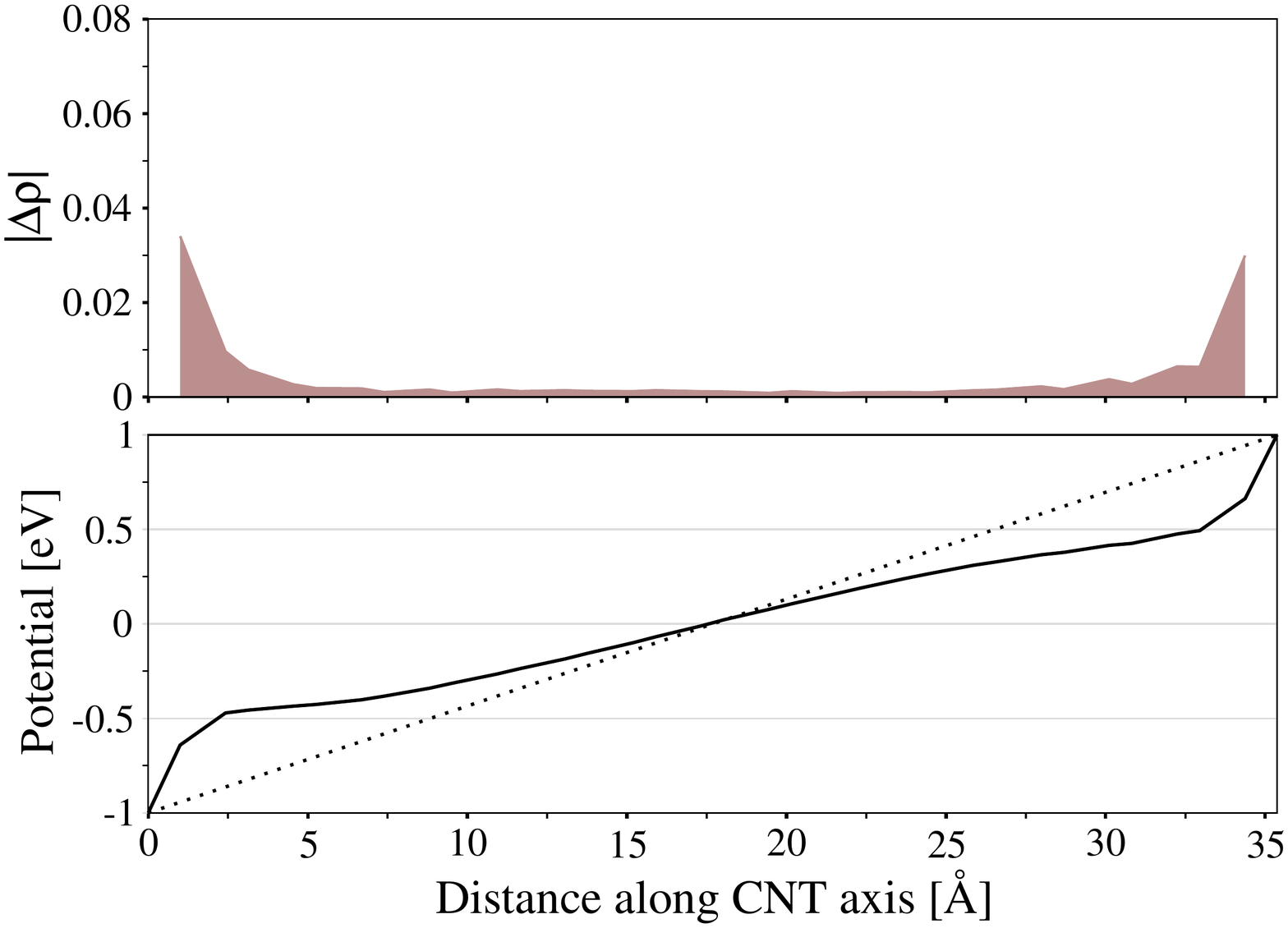}
	\caption{(upper) Absolute value of the change of electronic density $|\Delta\rho|$ for an 8-unit-cell (8,0) CNT under a 2V applied bias, as a function of the distance along the metal-CNT-metal device axis. (lower) Self-consistent potential $U_{SC}$ (line) and Laplace potential $U_{Laplace}$ (dots) for an 8-unit-cell (8,0) CNT under a 2V applied bias, as a function of the distance along the metal-CNT-metal device axis.}
	\label{fig:figure6}
\end{figure}

\begin{figure}
	\centering
		\includegraphics[width=0.5\columnwidth]{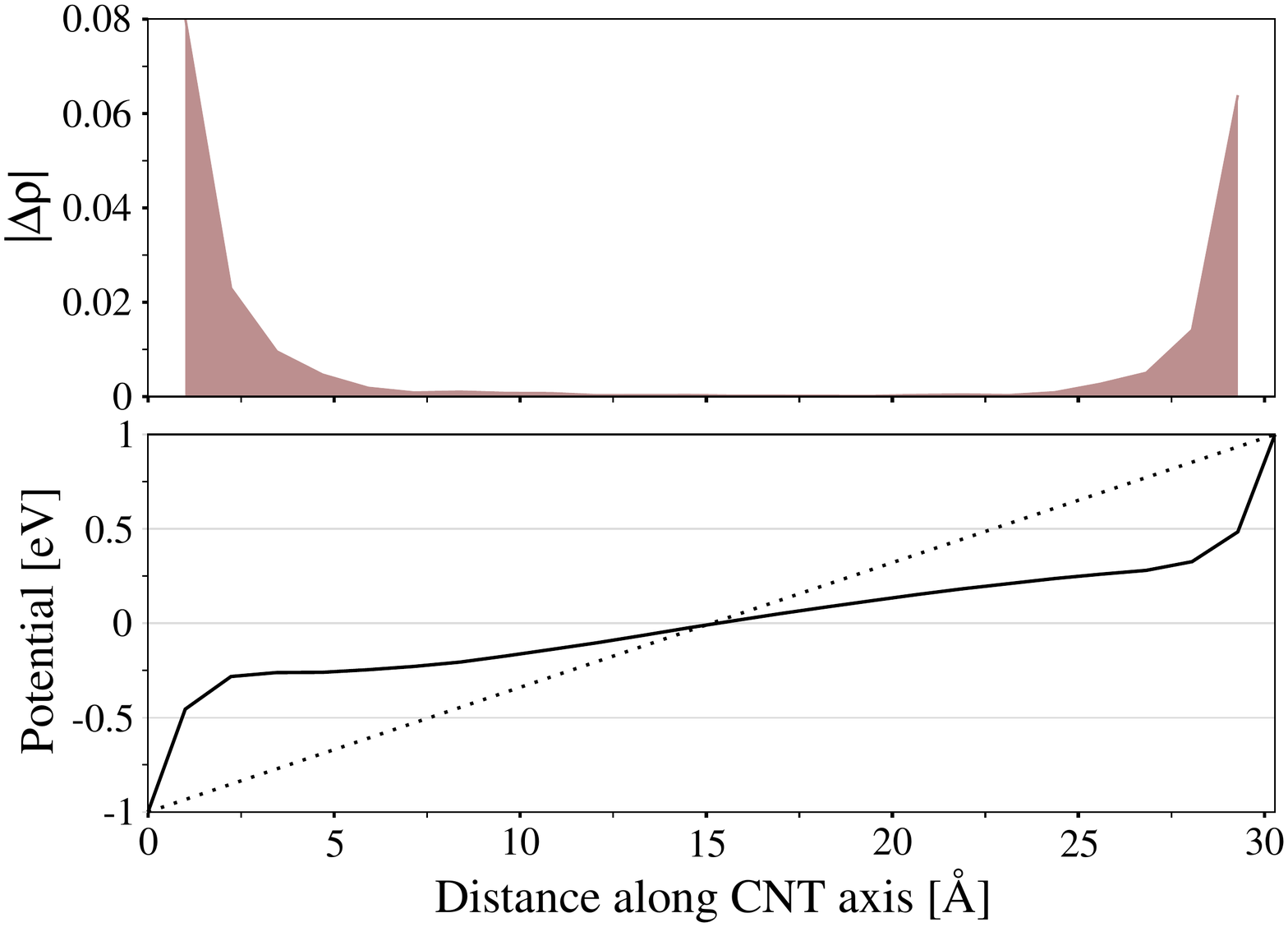}
	\caption{(upper) Absolute value of the change of electronic density $|\Delta\rho|$ for an 12-unit-cell (5,5) CNT under a 2V applied bias, as a function of the distance along the metal-CNT-metal device axis. (lower) Self-consistent potential $U_{SC}$ (line) and Laplace potential $U_{Laplace}$ (dots) for an 12-unit-cell (5,5) CNT under a 2V applied bias, as a function of the distance along the metal-CNT-metal device axis.}
	\label{fig:figure7}
\end{figure}

The investigation of molecular conduction in terms of local density of states only implicitly copes with the actual process of transport, which is the movement of charges. Principal quantum transport theory teaches that when the two electrodes of a molecular junction are bias-driven out of equilibrium, the electrochemical potentials of the latter separate and tend to pump in/pull out electrons from those molecular levels that lie within them\cite{Zahi03}. In the ideal case where the energy levels of the device are perfectly correlated to the contacts we can expect no accumulation of charges on the device body. This picture changes when dealing with real molecular systems where, as seen previously, the spatially anisotropic bidirectional movement of the LDOS's towards the energies of the electrochemical potentials can increase the number of localized states on the device (which could also exist prior to the application of bias due to symmetry/coupling factors). In this case we can expect a charge accumulation on the device body corresponding to the occupation of localized states that can condition the transport mechanism by reducing the easiness with which new charges can flow in/out of the device. This phenomenon can be visualized at figures \ref{fig:figure6} and \ref{fig:figure7}, where the absolute value of the change of electronic density between equilibrium and non-equilibrium conditions along the device body ($\Delta\rho=\rho-\rho_{eq}$) is plotted for the (8,0) and (5,5) CNTs for a 2V bias, accompanied by the respective self-consistent potential profile. Both electronic densities and potential values (interrelated via equation \ref{Usc}) are calculated on the atomic sites, whereas in the case of the potential profile mean values have been assigned for atoms with an identical projection on the device axis (atoms that belong to the same CNT ring). There are two distinct CNT areas where the change of electronic density has distinguishable characteristics, a) the area close to the metallic contacts and b) the inner part of the nanotube. In the first case, an initial screening can be observed for both types of CNTs (up to about 5\AA{} distance from the CNT ends), which could be perceived as the adjustment of the infinite metallic propagation modes to the finite modes of the molecular device. Such screening is quantitatively higher for the (5,5) tube and has an steep impact on the shaping of the potential profile near the CNT edges, contrary to the (8,0) case, where the deviation from the linear Laplace component of a neutral system is smaller (see fig. \ref{fig:figure6}). On the other hand, regarding the inner CNT body, a finite difference density distribution can be mainly observed for the semiconducting (8,0) tube, while minimal differences are evidenced for the (5,5) metallic one. This effect is directly related to the increase of localized states for the semiconducting CNT, which provokes an encapsulation of charges on the device body.

\section{Discussion}
In this article the influence of nonequilibrium charge transfer (NECT) on the conduction mechanism of molecular devices (CNTs, alkane dithiols) has been analyzed on the basis of local density of states alterations provoked by the application of bias. The main characteristic of this process can be individualized on the interaction between the molecular LDOS and the electrochemical potentials of the contacts, where a tracking relationship has been clearly manifested. Such process provokes a nonuniform reorganization of the density of states throughout the body of the molecular device, which, especially in geometrically symmetric devices with a conduction gap, can increase the number of localized states at energies close to the Fermi level. These states, when occupied by charges, can give rise to localized electronic wavefunctions that yield a non-zero difference between equilibrium and non-equilibrium electronic densities at the inner parts of the molecule. On the other hand it has been demonstrated that the extent of such effect is smaller in molecular systems with a metallic character.

It can be expected that the presence of the aforementioned processes is limited to the quantum description of electronic transport. Indeed, moving towards the semiclassical limit, the device can be thought of as the sum of numerous nanosize discretized units where quantum rules can be implemented. In this case, the potential profile that each of this units perceives is flat and therefore no bidirectional LDOS modifications are expected, rather than a unified movement of all states towards the respective potential level. Therefore no alteration of the local electronic structure should be observed with respect to the zero-bias case other than this shift.

The concepts discussed above can reflect both practically and theoretically on the perception of transport in molecular systems. From an engineering point of view, findings could be useful in the understanding of nanoscale device operation, since the dependence of the conduction gap formation from the applied source-drain bias magnitude is something that should not be overlooked. From a theoretical point of view, the presence of NECT phenomena can invoke a discussion both in a conceptual and in methodological level. Conceptually it becomes evident that molecular conduction issues cannot be solely treated in equilibrium terms. Methodologically it is confirmed that a self-consistent approach to the computer-aided design of molecular systems is fundamental for the capturing of charging phenomena, either in or out of equilibrium.

\begin{acknowledgments}
This work has been partially supported by the Sicilian Region under the contract POR-Regione Sicilia-Misura 3.15.
\end{acknowledgments}

\newpage

\end{document}